\documentclass[11pt,a4paper]{article}
\usepackage{jheppub} 

\usepackage{young}

\newcommand{\plaq}[1]{U_{{\rm P},#1}}  
\def\eq#1\en{\begin{equation}#1\end{equation}}  
\def\eqa#1\ena{\begin{align}#1\end{align}}
\def\eqg#1\eng{\begin{gather}#1\end{gather}}

\newcommand{\sumtwo}[2]%
{\mathop{\sum_{#1}}_{#2}}
\newcommand{\prodtwo}[2]%
{\mathop{\prod_{#1}}_{#2}}
\newcommand{\mintwo}[2]%
{\mathop{\min_{#1}}_{#2}}

\newcommand{\bV}{\bar{V}}

\newcommand{\nn}{\nonumber}
\def\dfrac#1#2{\displaystyle\frac{#1}{#2}}

\newcommand{\pslash}{p\kern-1ex /}
\newcommand{\qslash}{q\kern-1ex /}
\newcommand{\lslash}{l\kern-1ex /}
\newcommand{\sslash}{s\kern-1ex /}
\newcommand{\kaslash}{k_a\kern-2ex /}
\newcommand{\kbslash}{k_b\kern-2ex /}
\newcommand{\Dslash}{{\cal D}\kern-1.5ex /}

\newcommand{\tr}{{\rm tr}}

\newcommand{\beqa}{\begin{eqnarray}}
\newcommand{\eeqa}{\end{eqnarray}}

\def\tr{\text{tr}}

\newcommand{\be}{\begin{equation}}
\newcommand{\ba}{\begin{eqnarray}}
\newcommand{\ea}{\end{eqnarray}}
\newcommand{\ee}{\end{equation}}

\title{Entanglement entropy for pure  gauge theories in 1+1 dimensions using the lattice regularization}

\author[a,b]{Sinya Aoki,}
\author[c]{Etsuko Itou,}
\author[c]{Keitaro Nagata}

\affiliation[a]{Center for Gravitational Physics, Yukawa Institute for Theoretical Physics, Kyoto University, \\
Kitashirakawa Oiwakechou, Sakyo-ku, Kyoto 606-8502, Japan}
\affiliation[b]{Center for Computational Sciences, University of Tsukuba, Tsukuba 305-8577, Japan}
\affiliation[c]{KEK Theory Center, Institute of Particle and Nuclear Physics, High Energy Research Organization (KEK), Ibaraki 305-0801, Japan}

\emailAdd{saoki@yukawa.kyoto-u.ac.jp}
\emailAdd{eitou@post.kek.jp}
\emailAdd{knagata@post.kek.jp}

\abstract{
We study the entanglement entropy (EE) for pure gauge theories in 1+1 dimensions with the lattice regularization.
Using the definition of  the EE for lattice gauge theories proposed in a previous paper~\cite{Aoki:2015bsa},
we calculate the EE for arbitrary pure as well as mixed states in terms of eigenstates of the transfer matrix in 1+1 dimensional lattice gauge theory. 
We find that the EE of an arbitrary pure state does not depend on the lattice spacing, thus  giving the EE  in the continuum limit, and show  that the EE for an arbitrary pure state 
is independent of the real (Minkowski) time evolution.
We also explicitly demonstrate the dependence of EE on the gauge fixing at the boundaries between two subspaces,
which was pointed out for general cases in the paper~\cite{Aoki:2015bsa}.
In addition,
we calculate the EE at zero as well as finite temperature by the replica method,
and show that our result  in the continuum limit corresponds to the result  obtained before in the continuum theory,
with a specific  value of the counter term, which is otherwise arbitrary in the continuum calculation.
We confirm the gauge dependence of the EE also for the replica method.
}

\keywords{Entanglement entropy, pure gauge theories, lattice regularization, transfer matrix, replica method}

\subheader{\hfill YITP-16-96, KEK-CP-347, KEK-TH-1922}

\begin{document} 

\maketitle

\section{Introduction}
\label{intro}
The entanglement entropy (EE),  which tells us quantum properties of  a given state,
 plays important roles in many fields of physics including quantum field theories~\cite{Calabrese:2004eu, Calabrese:2009qy},
the string theory~\cite{t4, t1, sw1, sw2, r, t2, t3, t5}, 
condensed matter physics~\cite{wen, pre, mul, hal, wen2, gon2} and the physics of the black hole~\cite{sus, kab1, kab2, sh1, sh2}. 
For example, the EE is thought to be a useful tool to investigate confinement/deconfinement phase transition in gauge theories, in the contexts of the Gauge/Gravity correspondence~\cite{NT, dcp, lv} as well as a purely field theoretical approach~\cite{Nakagawa:2009jk, Nakagawa:2011su, Itou:2015cyu}.
  
To calculate the EE of a region $V$, one needs to express the whole Hilbert space
as a tensor product of the Hilbert space of $V$ 
and that of the complement of $V$ (denoted as $\bV$).
Due to the local gauge invariance in gauge theories, however,
the gauge invariant Hilbert space, which is characterized by gauge invariant operators
such as Wilson loops, 
cannot be decomposed into a tensor product of the gauge invariant subspaces of $V$ and $\bV$.
Because of this problem,
there is no unique way to define the EE in gauge theories~\cite{CHR2013, Radicevic2014, BP2008, Donnelly2012, Don}. 

In the previous paper~\cite{Aoki:2015bsa}, one of the present authors (S.A.) with collaborators  proposed a definition of the entanglement entropy in lattice gauge theories. 
They simply extended the gauge invariant Hilbert space to the whole Hilbert space of the link variables,
which is then decomposed into a tensor product of the Hilbert 
spaces of the link variables in the region $V$
and those in the region $\bV$.
Using  this decomposition, the EE can be defined  for an arbitrary subset of links,
and the definition can be applied to both  abelian and non-abelian gauge theories. 
They also discussed the issue of gauge invariance and pointed out the EE depends on the gauge fixing at the boundaries between $V$ and $\bV$. They applied their definition of the EE to $Z_N$ lattice gauge theories  and investigated  several properties including the gauge dependence and the EE of the topological states in arbitrary dimensions.
Similar proposals have been also made in refs.~\cite{Ghosh:2015iwa,Hung:2015fla, Soni:2015yga}.

In this paper, using the definition of the EE in ref.~\cite{Aoki:2015bsa}, we further study the EE for 
gauge theories based on the lattice regularization, 
which is nonperturbative and gauge invariant.
As a simple but non-trivial model, we consider pure gauge theories in 1+1 dimensions with the lattice regularization, and show that  the EE can be calculated analytically
using  the eigenstate of the transfer matrix in 1+1 dimensional lattice gauge theories.
We find that the EE  does not depend on both lattice spacing and lattice size, so that  the EE for an arbitrary pure state in the continuum limit can be 
automatically obtained. 
We also investigate the issue of the gauge dependence explicitly calculating the EE with various choices of the gauge fixing, and demonstrate that the EE depends on the gauge fixing if gauge transformations at boundaries between two subspaces are employed.

The present paper is organized as follows. 
In section~\ref{sec:LGT_2d}  we give a brief summary of lattice gauge theories in 1+1 dimensions such as the character expansion and the transfer matrix.
In section~\ref{sec:Operator}, we calculate the EE for arbitrary pure as well as mixed states in the lattice  gauge theories in 1+1 dimensions, using gauge invariant states in the Hilbert space (called the operator method), 
where any states can be expanded in terms of eigenstates of the transfer matrix. 
We show that the EE of an arbitrary pure state 
 does not depend on the lattice spacing, so that the continuum limit is automatically realized. In addition, the EE for an arbitrary pure state is found to be time independent. 
We also demonstrate that the EE depends on the gauge fixing at the boundaries,
as pointed out in the previous paper~\cite{Aoki:2015bsa}, by explicitly calculating the EE with various gauge fixings.
In section~\ref{sec:conclusion}, we conclude the paper.
In appendix~\ref{sec:Replica},  we calculate the EE for pure gauge theories in 1+1 dimensions on the lattice, using the replica method, at zero as well as finite temperatures.
We show that the EE in the continuum limit agrees with the EE in the continuum theory. Furthermore,  the EE obtained from lattice gauge theories  determines a value of the counter term, which cannot be fixed in the continuum calculation,
showing  one of the advantages of the lattice regularization. 
We finally confirm that the EE depends on the  gauge fixing at the boundaries also in this case.

\section{Lattice gauge theories in 1+1 dimensions}
\label{sec:LGT_2d}
In this section,  we collect several formula for lattice gauge theories in 1+1 dimensions, which will be used for the latter sections.

\subsection{Basic formula}
We denote $U(R)$ the (unitary) irreducible representation $R$ of $U \in G$ for a group $G$.
The plaquette action for gauge theories with the gauge group $G$ is given by
\beqa
S_p &=& \beta \sum_{n\in \mathbb{Z}^2} \tr  \left( \plaq{n}(F)+\plaq{n}^\dagger(F)-{\bf 2}\right) ,
\label{eq:plaquette}
\eeqa 
where the plaquette $\plaq{n}$ is defined as
\beqa
\plaq{n} &=& U_{n,1} U_{n+\hat 1,2} U_{n+\hat 2,1}^\dagger U_{n,2}^\dagger,
\eeqa
$F$ represents the fundamental representation (for example, $U(F)$ is an $N\times N$ unitary matrix for $G=$ SU(N) or U(N)), 
and $\hat \mu$ is the unit vector in the $\mu$ direction ($\mu=1,2$).
The inverse gauge coupling $\beta$ is related to the gauge coupling $g$
\beqa
\beta =\frac{1}{g^2 a^2},
\eeqa
where $g$ has the mass dimension one and $a$ is the lattice spacing.

The character expansion of each plaquette is given by
\beqa
\exp\left[ \beta\, \chi_F \left( \plaq{n}+\plaq{n}^\dagger-{\bf 2}\right) \right] &=&
\sum_R d_R \lambda_R(\beta) \chi_R( \plaq{n}), 
\eeqa
where 
\beqa
\chi_R(U) &=& {\rm tr}\, U(R), \quad d_R =\chi_R({\bf 1} ), \\
\lambda_R(\beta) &=& \frac{1}{d_R} \int dU\, \chi_R(U) \exp \left[ \beta \chi_F\left( U+U^\dagger-{\bf 2}\right) \right] .
\eeqa
This definition leads to
\beqa
0 \le \lambda_R(\beta) \le 1, \qquad \lambda_R(\beta)=1 \Leftrightarrow \beta=\infty.
\eeqa
For example, for the spin $j$ representation of $G=$SU(2), we have
\beqa
\lambda_j(\beta) &=& \frac{e^{-4\beta} I_{2j+1}(4\beta)}{4\beta} 
\eeqa
for  half integer $j$, where $I_n$ is the modified Bessel function. 
Formula of $\lambda_R(\beta)$ for other gauge groups  can be found in ref.~\cite{Drouffe:1983fv}. 

We give two important formula,
\beqa
\int d\Omega\, \chi_R ( A\Omega)\chi_{R^\prime}(\Omega^\dagger B) &=& \frac{1}{d_R} \delta_{RR^\prime} \chi_R(AB), 
\label{eq:basic1}\\
\int d\Omega\, \chi_R(A\Omega B\Omega^\dagger) &=&\frac{1}{d_R} \chi_R(A) \chi_R(B),
\label{eq:basic2}
\eeqa
which follow from
\beqa
\int d\Omega\, \Omega_{ab}(R) \Omega_{cd}^\dagger(R^\prime) &=& \frac{1}{d_R} \delta_{RR^\prime}\delta_{ad}\delta_{bc} .
\eeqa

\subsection{Transfer matrix, eigenvalues and eigenstates}
The transfer matrix $\hat T$ for the plaquette action (\ref{eq:plaquette}) on 1+1 dimensional lattice is given in ref.~\cite{Aoki:2015bsa} as
\beqa
T(U,V)\equiv \langle U \vert \hat T \vert V \rangle &=& \prod_{x=0}^{L-1} \exp \left\{\beta  \tr \left[ \left(U_x V_x^\dagger + V_x U_x^\dagger-\bf{2}\right) \right]\right\}
\eeqa
where $U_x, V_x$ are spatial link variables and $L$ is the number of links in  1-dimensional lattice. 
Using the character expansion, we write
\beqa
T(U,V) &=& \prod_{x=0}^{L-1} \sum_R d_R \lambda_R(\beta) \chi_R( U_x V_x^\dagger) .
\eeqa

An eigenfunction of the transfer matrix is easily obtained with the periodic boundary condition (PBC) as  
\beqa
R (U) &\equiv & \langle U \vert  R\rangle = \chi_R\left(  U \right),
\qquad U\equiv  \prod_{x=0}^{L-1} U_x, 
\eeqa
which satisfies 
\beqa
\langle U \vert \hat T \vert R \rangle &=& \int {\cal D} V\, T(U,V) R(V) =\prod_{x=0}^{L-1}
\int dV_x \, \sum_{R_x} d_{R_x}  \lambda_{R_x}(\beta)\chi_{R_x}( U_x V_x^\dagger)\chi_R(V) \nn \\
&=& \lambda_R^L(\beta)   \chi_R\left(\prod_{x=0}^{L-1}U_x\right) = \lambda_R^L(\beta) \langle U \vert R\rangle ,
\eeqa
so that the eigenvalue is $  \lambda_R^L(\beta)$, and has the correct normalization as 
\beqa
\langle R^\prime \vert R \rangle &=& \int \prod_{x=0}^{L-1}  dU_x\, \chi_{R^\prime} (U^\dagger) \chi_R(U)
=\delta_{R^\prime R} .
\eeqa

To understand the nature of the state $\vert R \rangle$, let us consider $G=U(1)$ case, where $R=n$ is an integer.
The positive (negative) $n$ represents how many times the Wilson line warps around the circle, the 1-dimensional space with the PBC, in the positive (negative) direction.

\section{Direct calculations of EE in lattice gauge theories}
\label{sec:Operator}
In this section, we directly calculate the EE from the operator method of the 1+1 dimensional lattice gauge theory.

\subsection{Density matrix and entanglement entropy}
We first consider the density matrix for an eigenstate $\vert R \rangle$ as
\beqa
\rho(R) &=& \vert R \rangle \langle R \vert, 
\eeqa
where we can take not only the ground state  $\vert 0 \rangle$ but also excited states $\vert R\not= 0\rangle$.
We here assume the PBC in space ({\it i.e.} 1-dimensional torus).

We take regions $A$ and $\bar A$ as
\beqa
A=\bigcup_{i=1}^\ell A_i, \qquad \bar A=  \bigcup_{i=1}^\ell B_i, 
\eeqa
where $A_i\cap A_j =\emptyset$ and  $B_i \cap B_j =\emptyset $ for $i\not= j$, and $A\cap \bar A=\emptyset$ by definition.  

The reduced density matrix is defined as
\beqa
\rho_A(R) &=& \tr_{{\cal H}_{\bar A}} \vert R\rangle \langle R \vert ,
\eeqa
which is explicitly given by
\beqa
 {}_A\langle U \vert \rho_A(R) \vert V \rangle_A &=& \prod_{i=1}^\ell \int dW_i  \chi_R( U_1W_1\cdots U_\ell W_\ell)
\chi_R(W_\ell^\dagger V^\dagger_\ell \cdots W_1^\dagger V_1^\dagger) \nn \\
&=&d_R^{-\ell} \prod_{i=1}^\ell \chi_R( U_i V_i^\dagger ),
\eeqa
where $U_i$ or $V_i$ is the ordered product of links in $A_i$, while $W_i$ is that in $B_i$. 
Using this, we can show
\beqa
{}_A\langle U \vert \rho_A(R)^2 \vert V \rangle_A &=& d_R^{-3\ell} \prod_{i=1}^\ell \chi_R( U_i V_i^\dagger )
= d_R^{-2\ell}  {}_A\langle U \vert \rho_A(R) \vert V \rangle_A,
\eeqa 
which means 
\beqa
\rho_A^2(R) &=& d_R^{-2\ell} \rho_A(R), \qquad \rho_A^n(R) = d_R^{-2(n-1)\ell} \rho_A(R),
\eeqa
so that 
\beqa
\tr_{{\cal H}_A}\, \rho_A^n(R) =  d_R^{-2(n-1)\ell} \tr_{{\cal H}_A}\, \rho_A(R) = d_R^{-2(n-1)\ell} . 
\eeqa
Therefore, the EE is obtained as
\beqa
S_R(A) &=& - \tr_{{\cal H}_A}\left[ \rho_A \log \rho_A\right] =
\lim_{n\rightarrow 1} \frac{1}{1-n}\log \left\{\tr_{{\cal H}_A}\, \rho_A^n(R)\right\} = 2\ell \log d_R,
\label{eq:EE}
\eeqa
showing that the EE is zero for the vacuum state ($R=0$) as $d_0=1$.
Remarkably, the EE  depends on neither the lattice spacing $a$ nor the number of lattice points $L$, so that
the EE in the continuum limit is nothing but eq.~(\ref{eq:EE}).
Furthermore, since the EE in eq.~(\ref{eq:EE}) only depends on $\ell$, the number of disjoint components of $A$,
the state $\vert R\rangle$ is regarded as the  topological state, where the EE is insensitive to both position and size of $A$.
Since the EE does not depend on the size of $A$, the strong subadditivity is trivially satisfied.
Note that the appearance of $\log d_R$ contributions to the EE in non-abelian gauge theories has been pointed out in refs.~\cite{Aoki:2015bsa,Soni:2015yga}.

We next consider a state consisting of a linear combination of eigenstates as
\beqa
\vert \left\{ c_R \right\} \rangle &\equiv& \sum_R c_R \vert R \rangle, \qquad \sum_R \vert c_R\vert^2 = 1,
\eeqa
whose density matrix is written as
\beqa
\rho(\left\{ c_R \right\} ) &=& \sum_{R,R^\prime} c_R \bar c_{R^\prime} \vert R \rangle \langle R^\prime \vert .
\eeqa
The reduced density matrix is thus given by 
\beqa
\rho_A (\left\{ c_R \right\}  ) &=& \sum_R \vert c_R\vert^2 \rho_A(R) ,
\eeqa
so that
\beqa
\rho_A^n (\left\{ c_R \right\} ) = \sum_R \vert c_R\vert^{2n} d_R^{-2(n-1)\ell} \rho_A(R) .
\eeqa
We thus obtain
\beqa
\tr_{{\cal H}_A}\, \rho_A^n (\left\{ c_R \right\} ) &=&  \sum_R p_R^n d_R^{-2(n-1)\ell},
\qquad p_R\equiv \vert c_R\vert^2, \\
S_{\{c_R\}}(A) &=& 2\ell \sum_R p_R \log d_R -\sum_R p_R\log p_R .
\label{eq:EE_pure}
\eeqa
Again the EE in eq.~(\ref{eq:EE_pure}) is considered to be 
the result in the continuum limit,
since it does not depend on the lattice spacing $a$.
In ref.~\cite{Soni:2015yga}, these two contributions to the EE in non-abelian gauge theories are called classical. It was also argued that these two contributions cannot be extracted in dilution or distillation experiments which involve only Local Operations and Classical Communication (LOCC). See ref.~\cite{Soni:2015yga} for more details.

Let us consider the real-time dependence of the state $\vert R\rangle$, controlled by the Schr\"odinger equation as 
\beqa
\hat H \vert R, t \rangle &=& \frac{\partial }{i\partial t} \vert R, t \rangle, \qquad \vert R, 0\rangle =\vert R \rangle,
\qquad \hat H = -\frac{1}{a} \log \hat T ,
\eeqa
which can be solved as
\beqa
\vert R, t\rangle &=& e^{i E_R t } \vert R \rangle, \qquad    E_R =-\frac{L}{a} \log \lambda_R(\beta).
\eeqa
Therefore, the time dependence for the state $\vert \{ c_R\} \rangle$ is given by $\vert \{c_R(t)\} \rangle$ with 
\beqa
c_R(t) = c_R e^{i E_R t} ,
\eeqa 
which however gives 
\beqa
p_R(t) = \vert c_R(t)\vert^2 = \vert c_R\vert^2= p_R.
\eeqa
This means that the EE for this state is independent under the real-time evolution described 
by the hamiltonian not only in the continuum limit but also  on the lattice.
Since the state here is not a thermal state, this time independence of the EE is nontrivial and is a special feature of gauge theories in $1+1$ dimensions.

Finally we consider the density matrix for the general mixed states given by
\beqa
\rho\left(\left\{ p_R\right\}\right) &\equiv& \sum_R p_R \vert R \rangle \langle R \vert .
\eeqa
For example, for a thermal state at temperature $T_B$, we have
\beqa
p_R = \frac{e^{- E_R/T_B}}{\displaystyle \sum_R e^{- E_R/T_B}}= \frac{\lambda_R^{L/(a T_B )} }
{\displaystyle \sum_R \lambda_R^{L/(a T_B )} }, 
\qquad \sum_R p_R=1, \quad p_R\ge 0 .
\eeqa
The reduced density matrix for this state becomes
\beqa
\rho_A\left(\left\{ p_R\right\}\right) &=& \sum_R p_R\rho_A(R) ,
\eeqa
and therefore we have
\beqa
\rho_A^n\left(\left\{ p_R\right\}\right) &=& \sum_R p_R^n d_R^{-2(n-1)\ell} \rho_A(R) ,
\eeqa
which leads to the same EE as before:
\beqa
S_{\{p_R\}}(A) &=& 2\ell \sum_R p_R \log d_R -\sum_R p_R\log p_R ,
\eeqa
where the second term is equal to the von Neumann entropy originated from the mixed state as
\beqa
S_{\rm mix} &=& - \tr \left[ \rho\left(\left\{ p_R\right\}\right) \log \rho\left(\left\{ p_R\right\}\right) \right]= -\sum_R p_R\log p_R .
\eeqa
For the thermal state, we have
\beqa
S_{\{p_R\}}(A) &=& \log\sum_R \lambda_R^{L/(aT_B)}(\beta) +  \frac{\displaystyle \sum_R \lambda_R^{L/(aT_B)}(\beta)
\left(2\ell \log d_R -\dfrac{L}{aT_B} \log \lambda_R(\beta)\right)}{\displaystyle\sum_R \lambda_R^{L/(aT_B)}(\beta)}, ~~~
\label{eq:EE_thermal}
\eeqa
which depends on the lattice spacing $a$, the temperature $T_B$ and the number of lattice points  $L$.
As shown in the appendix, this result agrees with the one obtained from the replica method.
We will also see that this EE in the continuum limit reproduces the previous result obtained in the calculation of the continuum theory~\cite{Gromov:2014kia}.

\subsection{Gauge fixing and gauge invariance}
As discussed in ref.~\cite{Aoki:2015bsa}, the EE is gauge invariant in the sense that it does not depend on the gauge fixing as long as no gauge fixing is employed at the boundary points while
gauge transformations including those at boundary points can change the value of  the EE.   
We explicitly demonstrate these properties mainly for the state $\vert R \rangle$ below.

\begin{figure}[tbh]
\begin{center}
\scalebox{0.4}{\includegraphics{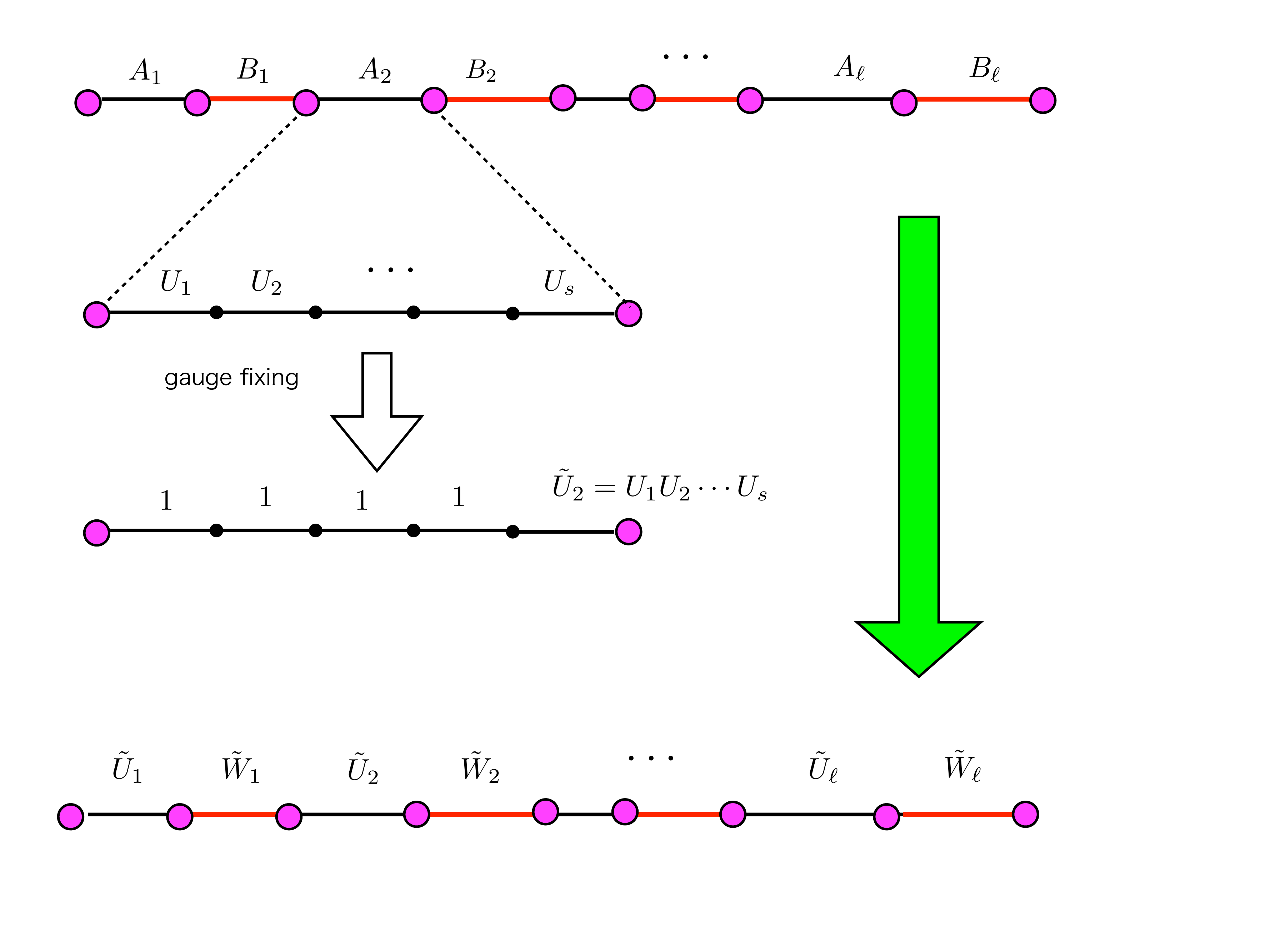}}
\end{center}
\vskip -1cm
\caption{The gauge fixing without gauge transformation at boundaries.
}
\label{fig:Gfix}
\end{figure}
We can set all link variables except one to an unit matrix in each $A_i$ or $B_i$, using gauge transformations at lattice points inside each region without boundary points so that $A_i \rightarrow \tilde U_i$ and $B_i\rightarrow \tilde W_i$
as shown in fig.~\ref{fig:Gfix}.
The corresponding reduced density matrix for the state $\vert R\rangle$ can be calculated as
\beqa
{}_A\langle \tilde U \vert \rho_A(R) \vert \tilde V \rangle_A &=& \prod_{i=1}^\ell \int d\tilde W_i  \chi_R( \tilde U_1\tilde W_1\cdots \tilde U_\ell \tilde W_\ell)
\chi_R(\tilde W_\ell^\dagger \tilde V^\dagger_\ell \cdots \tilde W_1^\dagger \tilde V_1^\dagger) \nn \\
&=&d_R^{-\ell} \prod_{i=1}^\ell \chi_R( \tilde U_i \tilde V_i^\dagger ),
\eeqa
which gives
\beqa
S_R(A) &=& 2\ell \log d_R .
\eeqa 
This shows that the EE after the gauge fixing remains the same as eq.~(\ref{eq:EE}) without gauge fixing.

We next consider the extreme case where gauge transformation at all points including all boundaries are used to fix the gauge. In this case, we can fix all link variables to an unit matrix except one (due to the PBC), which we take $\tilde U_1$ in $A_1$.  The reduced density matrix  is given as
\beqa
{}_A\langle \tilde U_1 \vert \rho_A(R)\vert \tilde V_1\rangle_A &=& \chi_R(\tilde U_1) \chi_R(\tilde V_1^\dagger) ,
\eeqa
which leads to $S_R(A) = 0$.

We finally consider more general case where gauge transformations including those at some boundary points are employed.
We fix link variables to an unit matrix except a few so that
non-trivial link variables are given in the following order
\beqa
\tilde U_1, \tilde W_1, \tilde U_2, \tilde W_2, \cdots \tilde U_s , \tilde W_s, \qquad \tilde U_i \in A,  \ \tilde W_i\in \bar A. 
\eeqa 
The corresponding reduced density matrix for the state $\vert R\rangle$ becomes
\beqa
{}_A\langle \tilde U \vert \rho_A(R) \vert \tilde V \rangle_A &=& \prod_{i=1}^s \int d\tilde W_i  \chi_R( \tilde U_1\tilde W_1\cdots \tilde U_s \tilde W_s)
\chi_R(\tilde W_s^\dagger \tilde V^\dagger_s \cdots \tilde W_1^\dagger \tilde V_1^\dagger) \nn \\
&=&d_R^{-s} \prod_{i=1}^s \chi_R( \tilde U_i \tilde V_i^\dagger ),
\eeqa
which gives
\beqa
S_R(A) &=& 2s \log d_R , \qquad s=1,2,\cdots, \ell.
\eeqa 
Here the order of $U$ and $W$ is important to obtain the above result.
For example, the order $\tilde  U_1, \tilde W_1, \tilde U_2, \tilde W_2$ corresponds to $s=2$, while
$ \tilde U_1,\tilde W_1,\tilde W_2, \tilde U_2$ to $s=1$ because of the PBC.

We thus conclude that  a possible value of
the EE for the state $\vert R\rangle$ is given as
\beqa
S_R(A) &=& 2s \log d_R , \qquad s=0,1,2,\cdots, \ell,
\eeqa
by some choice of the gauge transformations.

For the general state, it is easy to see that
\beqa
S_{\{p_R\}}(A) &=& 2s \sum_R p_R \log d_R -\sum_R p_R\log p_R ,\qquad s=0,1,2,\cdots, \ell,
\eeqa
where $p_R=\vert c_R\vert^2$ for the state $\vert \left\{ c_R\right\}\rangle $. 

\section{Conclusion}
\label{sec:conclusion}
In this paper, we calculate  the EE for the 1+1 dimensional pure gauge theories using the lattice regularization
with the operator method, and obtain
\beqa
S(A) &=&  \sum_R p_R \left( 2\ell \log d_R -\log p_R\right), \qquad p_R =\vert c_R\vert^2
\eeqa
for the state
\beqa
\vert \left\{ c_R\right\} \rangle &=& \sum_R c_R \vert R \rangle,
\eeqa
where $\vert R \rangle$ is the eigenstate of the transfer matrix and $R$ specifies the irreducible representation of the gauge group $G$. This result can be regarded as the continuum one as it  does not depend on the lattice spacing $a$.
A similar result is also obtained for the mixed states including the thermal state.

We explicitly confirm that the above EE can be reduced by the gauge transformation as
\beqa
S(A) &=&  \sum_R p_R \left( 2s \log d_R -\log p_R\right), \qquad s=0,1,\cdots, \ell,
\eeqa
as pointed out in ref.~\cite{Aoki:2015bsa}.

In appendix \ref{sec:Replica}, we calculate the same quantities using the replica method.
We will confirm that the results obtained by the replica method reproduce the EE for the vacuum state $\vert 0 \rangle$ as well as the thermal state in the main text.
We also confirm that the EE for the thermal state in the continuum limit reproduce the known continuum result~\cite{Gromov:2014kia}.
In addition, the value of the counter term can be fixed by the lattice calculation, 
contrary to the continuum treatment, which leave this term arbitrary~\cite{Gromov:2014kia}.
A similar  gauge dependence of the EE will be also demonstrated. 

\section*{Acknowledgement}
S. A. is supported in part by the Grant-in-Aid of the Japanese Ministry of Education, Sciences and Technology, Sports and Culture (MEXT) for Scientific Research (No. 16H03978) and by MEXT
and Joint Institute for Computational Fundamental Science (JICFuS)
as a priority issue ``Elucidation of the fundamental laws and evolution of the universe'' to be tackled by using Post K Computer. 
K.N. is supported by JSPS Grant-in-Aid for Scientific Research (No. 26800154 and 16H03988).
The authors thank the Yukawa Institute for Theoretical Physics at Kyoto University, where
this work was initiated during the YKIS2016 ``Quantum matter, Spacetime and Information''.

\appendix

\section{Entanglement Entropy  from the replica method}
\label{sec:Replica}
In this appendix, we calculate the EE for 1+1 dimensional lattice gauge theories using the replica method.

\subsection{Observables of lattice gauge theories in 1+1 dimensions } 
This subsection includes some useful formula of lattice gauge theories, which will be used in this appendix.

\begin{figure}[tbh]
\begin{center}
\scalebox{0.3}{\includegraphics{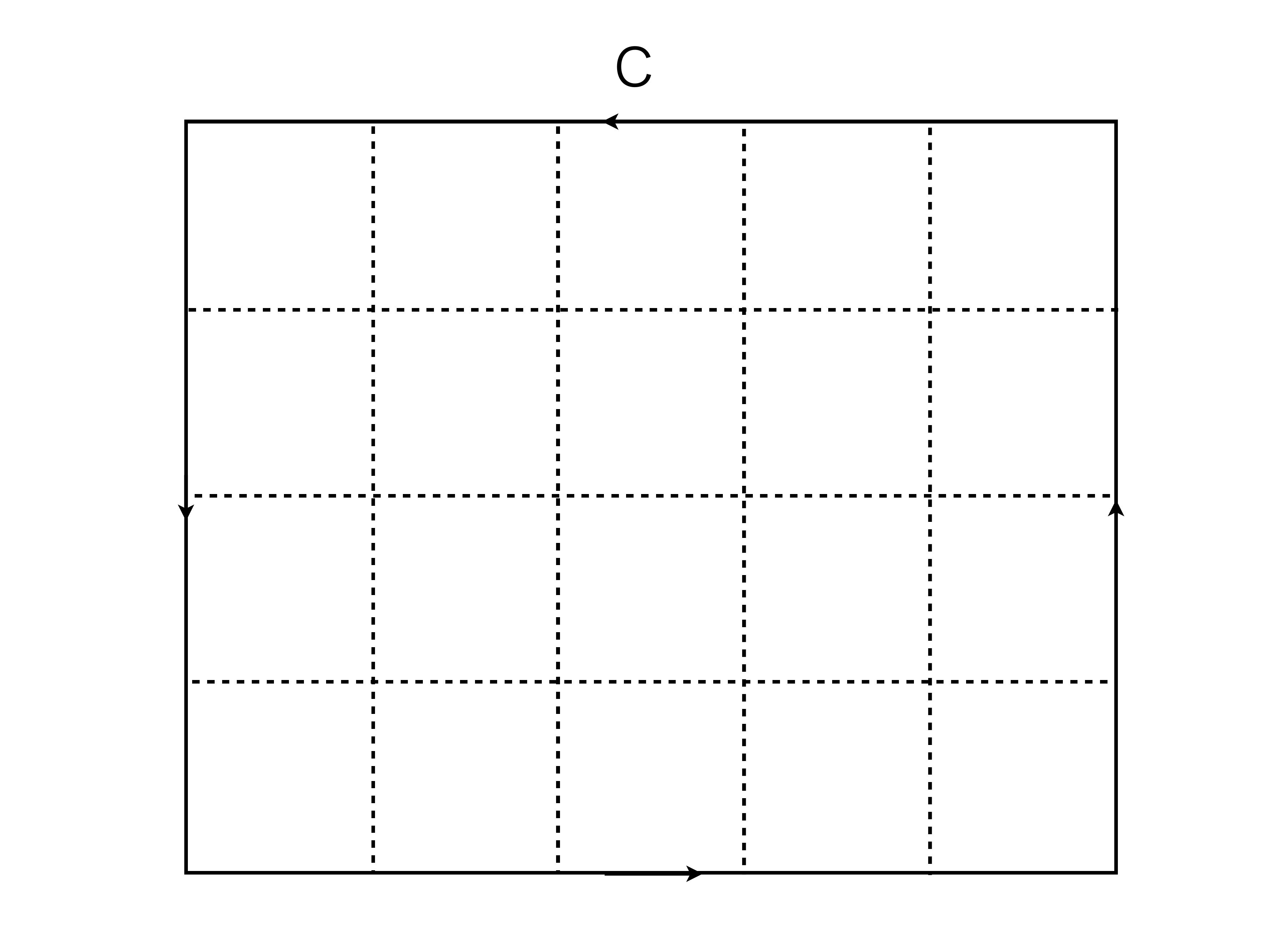}}
\end{center}
\caption{An example of a closed loop $C$. The dashed links belong to $C_{\rm in}$, while $\Gamma$ is the ordered product of links on $C$.  The action $S_p(C)$ consists of plaquettes inside $C$, ones made of dashed links only and 
the 
others made of both dashed and solid links. 
$\#C$, which is the number of the plaquettes, is $20$ in this loop $C$.
}
\label{fig:LoopC}
\end{figure}
Let us consider the quantity defined by
\beqa
K_C(\Gamma) &=& \int \prod_{\ell\in C_{\rm in}} dU_\ell \, e^{S_p(C)}, \qquad
\Gamma = \prod_{\ell\in C} U_\ell
\eeqa
for the closed loop $C$ (for example, see fig.~\ref{fig:LoopC}), and $C_{\rm in}$ represents the links inside $C$  (dashed links in fig.~\ref{fig:LoopC}) without links on $C$. 
Here $\ell =(n,\mu)$ represents a link between $n$ and $n+\hat\mu$
and $S_p(C)$ is the action consisting of the plaquettes inside the loop $C$.
Using the formula in eqs.~(\ref{eq:basic1}) and (\ref{eq:basic2}), we obtain
\beqa
K_C(\Gamma) &=& \sum_R d_R \lambda_R^{\# C}(\beta) \chi_R(\Gamma) ,
\label{eq:basic}
\eeqa
where $\# C$ is the number of plaquettes inside $C$.

Thus, the partition  function with the PBC in both directions is given by
\beqa
Z_{\rm PBC} &=& \sum_R \lambda_R^{LT}(\beta),
\eeqa
where $L T$ is the total number of space-time plaquettes. 

We next consider the expectation value of a $L_0\times T_0$ Wilson loop for the irreducible representation $R (\not=0)$ with  the PBC. Since the product of two irreducible representations is decomposed 
as
\beqa
R \otimes R_a = \bigoplus_b N_{R,R_a}^{R_b}  R_b ,
\eeqa
which means each irreducible representation $R_b$ appears $N_{R,R_a}^{R_b} $ times in the product of $R\otimes R_a$, we have
\beqa 
\langle \chi_{R} (U_{L_0\times T_0}) \rangle &=&\sum_{R_a,R_b}  \frac{d_{R_a} d_{R_b}}{Z_{\rm PBC}}
N_{R,R_a}^{R_b}  \lambda_{R_b}^{L_0T_0}(\beta) \lambda_{R_a}^{LT-L_0T_0}(\beta).
\eeqa
Assuming that $L_0T_0 \ll LT$ and taking the large $LT$ limit, the term with $R_a=0$  dominates in the above,
so that $R_b=R$ and $N_{R,0}^{R}=1$,  which leads to
\beqa
\langle \chi_{R} (U_{L_0\times T_0}) \rangle &\simeq& d_{R} \left(\frac{\lambda_{R}(\beta)}{\lambda_0(\beta)}\right)^{L_0T_0},
\eeqa
where we use the fact that $\lambda_0(\beta) > \lambda_R(\beta)$ ($R\not= 0)$ for all $\beta < \infty$.
Thus the static quark potential for the representation $R$ is given by
\beqa
V_R(L_0 ) a &=& -\lim_{T_0\rightarrow \infty} \frac{1}{T_0} \log  \langle \chi_{R} (U_{L_0\times T_0}) \rangle
= - L_0 \log  \left(\frac{\lambda_{R}(\beta)}{\lambda_0(\beta)}\right),
\eeqa
which increases linearly in $L_0$, showing the confinement. The string tension in the physical unit  is obtained as
\beqa
\sigma_{R}  a^2 &=& -  \log  \left(\frac{\lambda_{R}(\beta)}{\lambda_0(\beta)}\right).
\eeqa

\def\YGbox{8.0}
In the continuum limit that $\beta\rightarrow\infty$, we have
\beqa
\lim_{\beta\rightarrow\infty} \left(\frac{\lambda_R(\beta)}{\lambda_0(\beta)}\right) =\frac{1-c(R) \beta^{-1}}{1-c(0)\beta^{-1}} +O\left(\beta^{-2}\right) \simeq 1-g^2 a^2 C_2(R)
\eeqa
where $C_2(R) = c(R) - c(0)$. For SU(2), we have
\beqa
c(j)&=&(j+1/4)(j+3/4), \qquad C_2(j)=j(j+1)
\eeqa
for $j=0,1/2,1,3/2,2\cdots$. Note that $C_2(j)$ is the quadratic Casimir of the spin $j$ representation, and this is true in general that $C_2(R)$ is the quadratic Casimir of the irreducible representation $R$ of the group $G$, defined by
\beqa
\sum_a T^a(R) T^a(R) &=&C_2(R){\bf 1},
\eeqa
where $T^a(R)$ is the irreducible representation $R$ for the generator $T^a$ of the group $G$. 
For example, $C_2(n) = n^2$
for the $R=n\in \mathbb{Z}$ representation of U(1) group, while $C_2(q_1,q_2) =q_1+q_2 +(q_1^2+q_1 q_2 +q_2^2)/3$ for the representation $(q_1,q_2)$ of SU(3) group, where
$q_i = $ (number of boxes in row $i$) $-$ (number of boxes in row $(i+1)$) in the Young tableau of SU(N) group~\cite{Drouffe:1983fv}.  
Casimir invariants for  a few low-lying representations of SU(N) group can be found in ref.~\cite{SK1982} and are given in Table~\ref{tab:casimir}, where
$\YoungTab[0][\small]{{n}}$ represents $N-n$ boxes in a column.
\begin{table}[tbh]
\begin{center}
\begin{tabular}{|l|c|c|c|}
\hline
Rep. & $(q_1,q_2,\cdots,q_{N-1})$ & $d_R $ & $C_2(R)$ \\
\hline
\Young[0]{1} & $(1,0^{N-2})$ & $N$ & $(N^2-1)/(2N)$ \\
$\YoungTab[0]{{1}}$ & $(0^{N-2},1)$ & $N$ & $(N^2-1)(2N)$ \\
 $\YoungTab[0]{{1,}}$& $(1,0^{N-3},1)$ & $N^2-1$ & $N$ \\
 \Young[0]{2} & $(2,0^{N-2})$ & $N(N+1)/2$ & $(N-1)(N+2)/N$ \\
 \Young[0]{11} & $(0,1,0^{N-3})$ & $N(N-1)/2$ & $(N+1)(N-2)/N$ \\
 \Young[0]{3} & $(3,0^{N-2})$ & $N(N+1)(N+2)/6$ & $3(N-1)(N+3)/(2N)$ \\  
 \Young[0]{21} & $(1,1,0^{N-3})$ & $N(N^2-1)/3$ & $3(N^2-3)/(2N)$ \\
 \Young[-1]{111} & $(0,0,1,0^{N-4})$ & $N(N-1)(N-2)/6$ & $3(N+1)(N-3)/(2N)$ \\
$\YoungTab[0]{{1,,}}$ & $(2,0^{N-3},1)$ & $N(N+2)(N-1)/2$ & $(3N-1)(N+1)/(2N)$ \\
$\YoungTab[0]{{1}}\Young[-1]{11}$ & $(0,1,0^{N-4},1)$ & $N(N-2)(N+1)/2$ & $(3N+1)(N-1)/(2N)$ \\
$\YoungTab[0]{{1,1,}}$ & $(1,0^{N-3},2)$ & $N(N+2)(N-1)/2$ & $(3N-1)(N+1)/(2N)$ \\
$\YoungTab[0]{{2,}}$ & $(1,0^{N-4},1,0)$ & $N(N-2)(N+1)/2$ & $(3N+1)(N-1)/(2N)$ \\
\hline
\end{tabular}
\end{center}
\caption{Invariant Casimir for low-lying representations of SU(N) }
\label{tab:casimir}
\end{table}    
 
Using the formulas, the string tension for the irreducible representation $R$ is given by
\beqa
\sigma_R &=&C_2(R)g^2 
\eeqa
in the continuum limit.

\subsection{Replica method}
We calculate the EE 
for a $1+1$ dimensional lattice with spatial lattice points $L$ and temporal lattice points $T$ (see fig. \ref{fig:Replica}) using the formula
\beqa
S(A,LT) &=& -\lim_{n\rightarrow 1} \frac{\partial}{\partial n} \tr_{{\cal H}_A} \rho_A^n =   
-\lim_{n\rightarrow 1} \frac{\partial}{\partial n} \log ( \tr_{{\cal H}_A} \rho_A^n), \nonumber \\
& = & \lim_{n\rightarrow 1}  \frac{1}{1-n} \log\,  \tr_{{\cal H}_A}\, \rho_A^n,
\eeqa
where $\rho_A$ is the reduced density matrix $\rho_A = \tr_{{\cal H}_{\bar A}} \rho$, and ${\rm tr}\, \rho_A^n$ can be evaluated by the replica method as
\beqa
 {\rm tr}\, \rho_A^n &=& \frac{Z_n (LT)}{Z_1^n} , \qquad Z_1=Z_1(LT),
\eeqa
where $Z_1(LT) $ is the unnormalized partition function of the original theory.

As before, we consider the region $A$ and its compliment $\bar A$ in 1-dimension as the union of $\ell$ disjoint regions $A_i$ and $B_i$.  
The whole space can be expressed as $ (A_1, B_1,A_2,B_2,\cdots , A_\ell, B_\ell) $.

\subsection{Calculation}
\begin{figure}[tbh]
\begin{center}
\scalebox{0.4}{\includegraphics{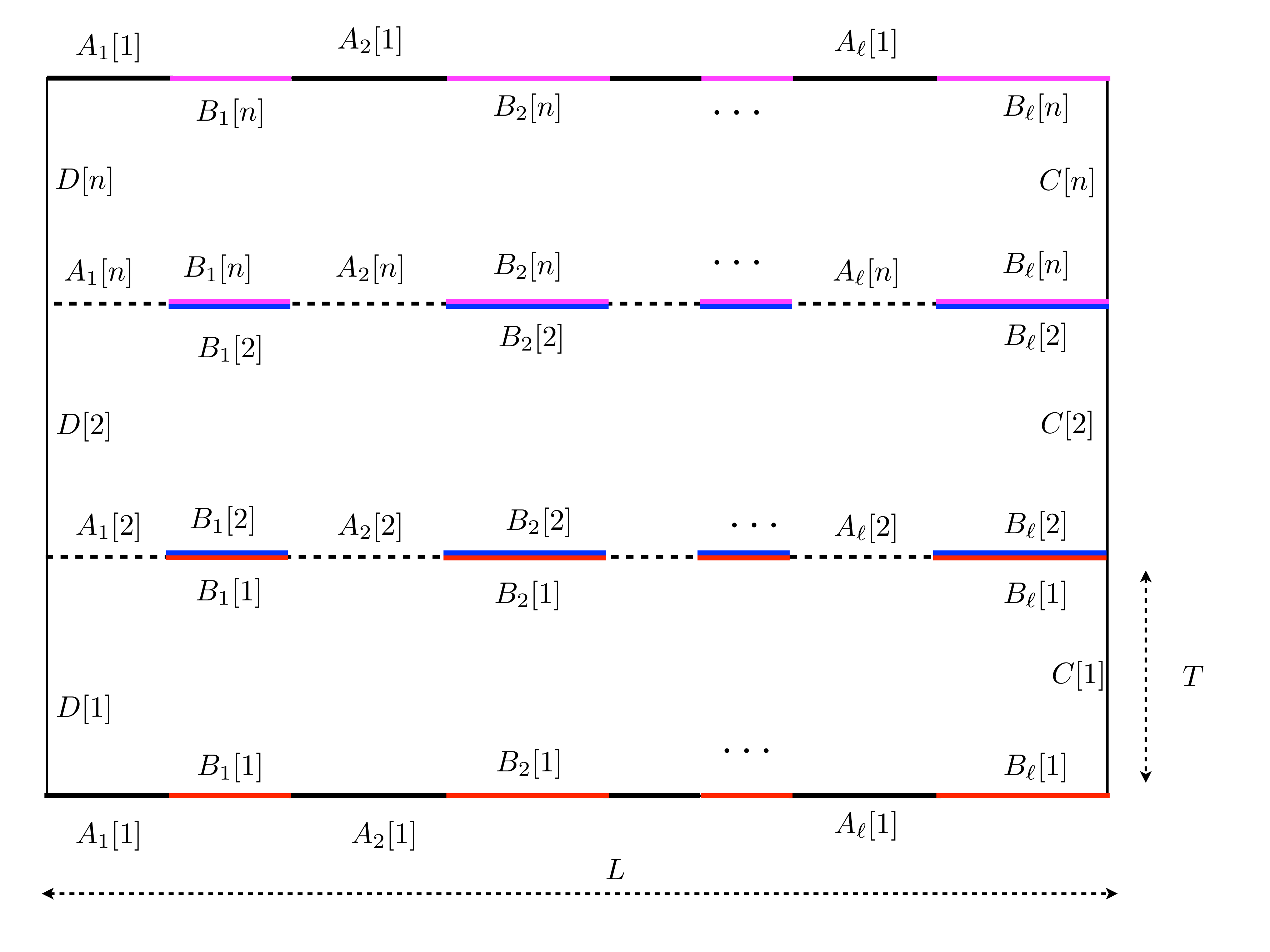}}
\end{center}
\caption{The replica configuration for $n=3$.
}
\label{fig:Replica}
\end{figure}

Let us calculate $Z_n(LT)$ 
using the character expansion. 
It is easy to see that eq.~(\ref{eq:basic}) leads to
\beqa
Z_n(LT) &=& \int{\cal D} U   \prod_{k=1}^n\sum_R d_R \lambda_R^{LT} (\beta) \chi_R\left(A_1[k] B_1[k] A_2[k] B_2[k] \cdots A_\ell[k] B_{\ell}[k] \right.\nn \\
&\times&  \left. C[k] B_{\ell}^\dagger[k] A_\ell^\dagger[k+1] \cdots  B_1^\dagger[k] A_1^\dagger[k+1] D^\dagger[k]
\right) \label{eq:def_replica} \\
& \equiv &  \int_{A\cup C\cup D}{\cal D} U  \prod_{k=1}^n Z_k(A) , 
\eeqa
where $A_i[k]$ and $B_i[k]$ represent the ordered products of the spatial links in the regions $A_i$ and $B_i$ in the $k$-the replica while $C[k]$ and $D[k]$ are the ordered products of the temporal links at spatial boundaries,
$x=L$ and $x=0$. $L T$ is the number of the plaquettes in one-replica.  
Here $\int {\cal D}U$ represents integrations of all links on $A,B,C,D$, while $\int_{A\cup C\cup D}{\cal D}U $ means integrations of links on $A,C,D$ only.
Since the region $A_i$ in $k$-th replica is connected to the same region in the $k+1$ replica, $A_i[k+1]^\dagger$ appears in the above formula, while the trace over $B_i[k]$ is implied within $k$-th replica. Note that $A_i[n+1] = A_i[1]$. 
See fig.~\ref{fig:Replica} for the $n=3$ case.

\begin{figure}[tbh]
\begin{center}
\scalebox{0.4}{\includegraphics{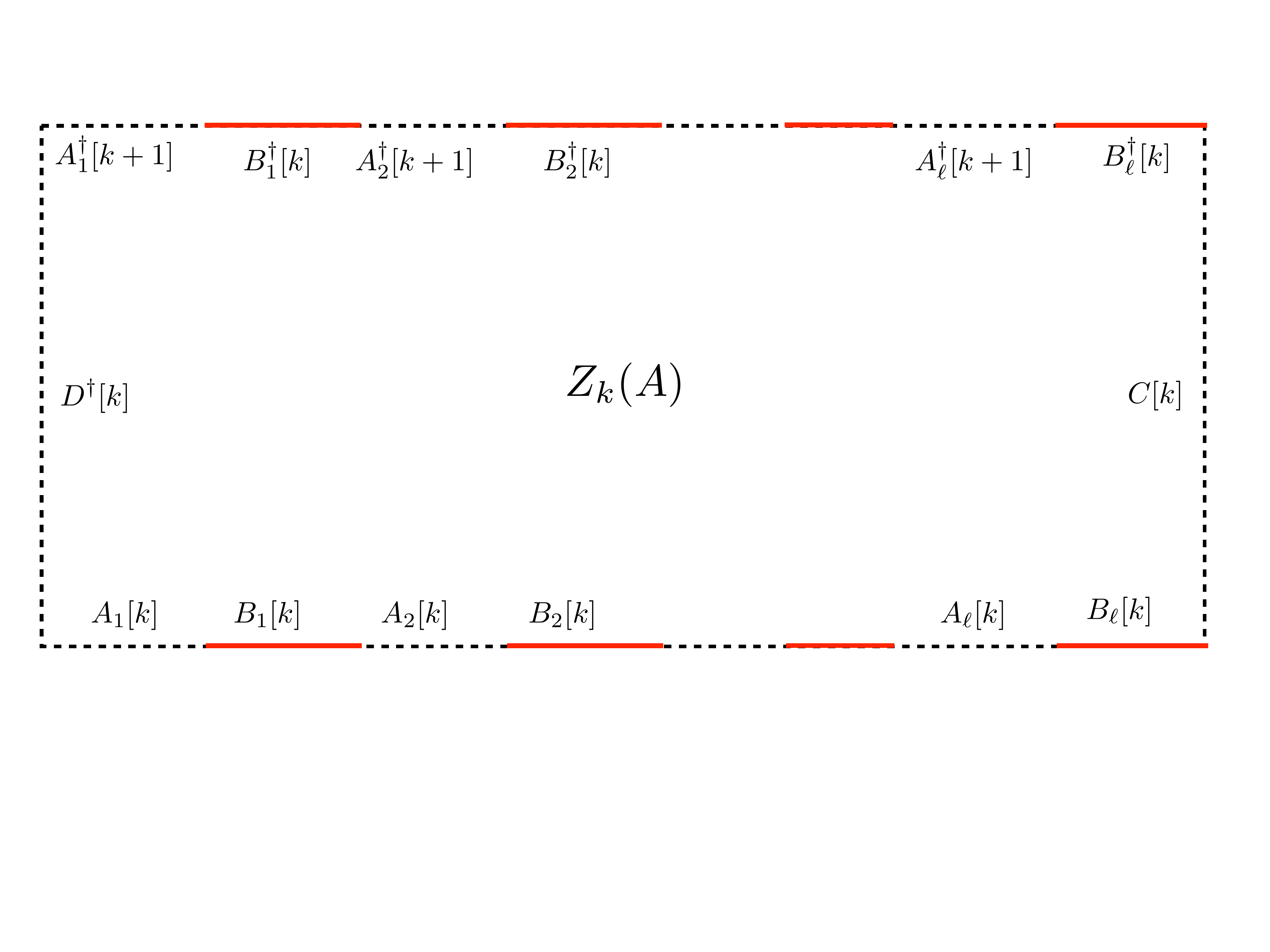}}
\end{center}
\vskip -3cm
\caption{The configuration of $Z_k(A)$.
}
\label{fig:Zk}
\end{figure}

For each $k$, we integrate over $\Omega_i$ of $B_i[k] = \Omega_i \hat B_i[k] $ using eq.~(\ref{eq:basic2}) (see fig.~\ref{fig:Zk}), and obtain
\beqa
 Z_k(A)  &=& \sum_R d_R^{1-\bar\ell} \lambda_R(\beta)^{LT}  
\chi_R(D[k]^\dagger A_1[k] A_1^\dagger [k+1]) 
\prod_{i=2}^\ell \chi_R(A_i[k] A_i^\dagger [k+1] )\nn \\
&\times&  \chi_R(C[k])\ .
\eeqa
We then integrate over $U_i$ of $A_i[k]= U_i \hat A_i[k]$ in $Z_n(A)$ using eq.~(\ref{eq:basic1}) as
\beqa
 \int dU_i \,  \chi_R (A_i[k-1] A_i^\dagger [k] ) \chi_{R^\prime} (A_i[k] A_i^\dagger [k+1] ) 
 &=&\delta_{RR^\prime} \frac{1}{d_R}   \chi_R( A_i[k-1] A_i^\dagger [k+1] )~~~ 
\eeqa
and 
\beqa
 &&\int dU_1\,  \chi_R (D[k-1]^\dagger A_1[k-1] A_1^\dagger [k] ) \chi_{R^\prime} (D[k]^\dagger A_1[k] A_1^\dagger [k+1] )  \nn \\
 &=&\delta_{RR^\prime} \frac{1}{d_R}    \chi_R( D[k]^\dagger D[k-1]^\dagger A_1[k-1] A_1^\dagger [k+1] ) ,
\eeqa
which lead to
\beqa
Z_n(A) &=&
 \sum_R \lambda_R^{n LT}(\beta) d_R^{-(n-1)(2\ell-1)} 
 \chi_R (\prod_{k^\prime=n}^1 D[k^\prime]^\dagger ) \prod_{k=1}^n \chi_R( C[k] ).  ~~~
\label{eq:Zn}
\eeqa

The periodic boundary condition implies the integration of eq.~(\ref{eq:Zn}) over $C=D$, which gives
\beqa
Z_n(A) &=& \sum_R \lambda_R^{n LT}(\beta) d_R^{-(n-1)2\ell }.
\label{eq:Zn_PBC}
\eeqa

Using the above results, we obtain
\beqa
{\rm tr}\, \rho_A^{n}  &=&\frac{ \sum_R \lambda_R^{n LT}(\beta) d_R^{-(n-1)2\ell}}{\left(\sum_R  \lambda_R^{ LT}(\beta)\right)^n}.
\eeqa

\subsection{Entanglement Entropy}
Taking $n\rightarrow 1$ limit, the entanglement entropy is given by
\beqa
S(A, LT) &=& \log \sum_R\lambda_R^{LT}(\beta) -\frac{\sum_R\lambda_R^{LT}(\beta)\log \lambda_R^{LT}(\beta)}{\sum_R \lambda_R^{LT}(\beta)}
+2\ell \frac{\sum_R \lambda_R^{LT}(\beta) \log d_R }{\sum_R \lambda_R^{LT}(\beta)},~~~
\eeqa
which shows that the EE does not depend on the size of the region $A$ but depend on the number of the boundaries of $A$, $2\ell$. This $S(A,LT)$ completely agrees with the EE in eq.~(\ref{eq:EE_thermal}) for the thermal state at the temperature $ T_B =1/(Ta) $ with the finite size $La$.
At the zero temperature that $1/T\rightarrow 0$ (or the thermodynamical limit that $L\rightarrow\infty$), the EE goes to zero as
\beqa
\lim_{1/(Ta)\rightarrow 0} S(A,LT) &=& 2\ell \log d_0 = 0, 
\eeqa
since $\lambda_0 (\beta) > \lambda_{R\not=0} (\beta)$ at $\beta <\infty$ and $d_0=1$.
This result also agrees with the one in the main text.

Even though the replica method correctly gives the EE for the vacuum state as well as the thermal state,  the operator method in the main text is much more powerful
to calculate the EE for 1+1 dimensional gauge theories as it can give the EE for an arbitrary state.

Let us consider the continuum limit. Since
\beqa
\lim_{\beta\rightarrow\infty}\frac{\lambda_R^{LT}(\beta)}{\lambda_0^{LT}(\beta)} &=& e^{-v_R}, \qquad
v_R =  v C_2(R), 
\eeqa
where $v = g^2 l_p t_p$ with $l_p =La$ and $t_p=Ta$ is the size of the 2-dimensional space-time in unit of $g^{-2}$, we obtain
\beqa
S(\ell, v) &=& \sum_{R\not=0} (v_R +2\ell \log d_R) \frac{ e^{-v_R}}{f(v)} + \log f(v),
\eeqa
where
\beqa
f(v) &=& 1+\sum_{R\not= 0} e^{-v_R} .
\eeqa
This result shows that the EE is finite in the continuum limit and agrees with the previous result, eq.~(18) of ref.~\cite{Gromov:2014kia},  
\beqa
S(A) &=& 2 \ell v + \ln\left(\sum_R e^{-\frac{1}{2T_BN} C_2(R)}\right) - \frac{\sum_R e^{-\frac{1}{2T_BN} C_2(R)}\ln d_R^{-2\ell} e^{-\frac{1}{2T_BN} C_2(R)}}{\sum_R e^{-\frac{1}{2T_BN} C_2(R)}},
\eeqa
obtained directly in the continuum theory, with the identification that  $1/(2T_BN) = g^2 l_p t_p$.
Note that our result corresponds to the specific value of the UV regularization parameter $v$, $v=0$,
 in eq.~(18) of ref.~\cite{Gromov:2014kia}.

\subsection{Gauge fixing and gauge invariance}
As discussed in the main text,
the EE depends on the gauge fixing if gauge transformations at boundary points are employed. 

We start with $Z_n(A)$ in eq.~(\ref{eq:def_replica}). 
We can set all spatial link variables to an unit matrix except one in each $A_i$ or $B_i$, using gauge transformations at lattice points inside each region without  boundary points.   
After this gauge fixing, $Z_n(A)$ is given by eq.~(\ref{eq:def_replica}) with the replacement that $A_i[k] \rightarrow \tilde U_i[k]$ and $B_i[k]\rightarrow \tilde V_i[k]$, where $\tilde U_i[k]$ or $\tilde V_i[k]$ a (non-gauge fixed) link variable in $A_i$ or $B_i$ of the $k$-th replica (fig.~\ref{fig:Gfix}), while $C[k]=D[k]$  are unchanged.
It is now clear that we obtain the same result, eq.~(\ref{eq:Zn_PBC}), after integrating out $U_i[k]$ and $V_i[k]$. The gauge invariance of $Z_n(A)$ leads to the gauge invariance of $S(A)$.

Using all gauge transformation including those on the boundaries, we can set all spatial link variables in $A\cup \bar A$  to an unit matrix except one.
We take non-trivial link variable  in $A_1$ for each replica and denote it as $U[k]$. 
We then have
\beqa
Z_n(A)&=& \prod_{k=1}^n Z_k(A)
\eeqa
where
\beqa
Z_k(A) &=& \Bigl\langle \sum_R d_R \lambda_R^{LT}(\beta)\chi_R(D^\dagger[k]  U[k] C[k] U[k+1] )\Bigr\rangle ,
\eeqa
which leads to
\beqa
Z_n(A) &=& \sum_R \lambda_R^{nLT}(\beta) d_R \langle \chi_R(\prod_{k=n}^1 D[k]^\dagger U[1] 
\prod_{k=1}^n C[k] U^\dagger[1])\rangle \nn \\
&=& \sum_R \lambda_R^{nLT}(\beta) \langle\chi_R(\prod_{k=n}^1 D[k]^\dagger) \chi_R(\prod_{k=1}^n C[k])\rangle =  \sum_R \lambda_R^{nLT}(\beta) .
\eeqa

For more general gauge fixings including some boundary points, there appear $s$  disjoint regions $\tilde A_i$ and $\tilde B_i$ with $i=1,2,\cdots, s$, each of which has at least one non-unity
link. Here we can have $ 1 \le s \le \ell$.
In this case, we have
\beqa
Z_{n} (A,s) &=& \sum_R \lambda_R^{nLT}(\beta) d_R^{-(n-1)2 s}.
\eeqa

Considering all cases, we have
\beqa
S(A,LT) &=&  \log \sum_R \lambda_R^{LT}(\beta) -\frac{\displaystyle\sum_R  \lambda_R^{LT}(\beta)  \log \lambda_R^{LT}(\beta) }{\displaystyle\sum_R  \lambda_R^{LT}(\beta) }
+2s \frac{\displaystyle\sum_R  \lambda_R^{LT}(\beta)  \log d_R }{\displaystyle\sum_R  \lambda_R^{LT}(\beta) }
\eeqa
where $s=0,1, 2,\cdots, \ell$, for the EE after some gauge fixing.

\providecommand{\href}[2]{#2}\begingroup\raggedright

\end{document}